\documentclass[12pt,12pt]{article}
\usepackage[numbers]{natbib}
\usepackage{amssymb}
\usepackage{amsmath}
\usepackage{enumerate}
\usepackage{graphicx}
\usepackage{graphics}
\usepackage{textcomp}
\usepackage{multirow}
\usepackage[T1]{fontenc}
\usepackage[utf8]{inputenc}
\usepackage{authblk}
\usepackage{xcolor}

\usepackage[normalem]{ulem}

\begin{document}

\begin{center}

{\Large \bf Holographic $k$-string Tensions in Higher Representations}

\vskip .3cm
{\Large \bf  and L\"uscher Term Universality}

\vskip 1.3 cm

{\large   B. Button$^1$, S. J. Lee$^1$, L. A. Pando Zayas$^2$, V. Rodgers$^1$  and K. Stiffler$^3$}

\vskip .4cm \centerline{\it ${}^1$ Department of Physics and Astronomy}
\centerline{\it The University of Iowa,}
\centerline{\it Iowa City, IA 52242}

\vskip .4cm \centerline{\it ${}^2$ Michigan Center for Theoretical
Physics}
\centerline{ \it Randall Laboratory of Physics, The University of
Michigan}
\centerline{\it Ann Arbor, MI 48109-1120}

\vskip .4cm \centerline{\it ${}^3$  Maryland Center for Fundamental Physics}
\centerline{ \it John S. Toll Physics Building,  The University of Maryland}
\centerline{\it College Park, MD 20742}

\end{center}

\vskip 1 cm


\begin{abstract}
We investigate a holographic description of $k$-strings in higher representations via D5 branes with worldvolume fluxes. The D-brane configurations are embedded in supergravity backgrounds dual to confining field theories in 3 and 4 dimensions. We compute the tensions and find qualitative agreement for the totally symmetric and totally anti-symmetric representations with the results of other methods such as lattice as well as the Hamiltonian approach of Karabali and Nair. A one-loop computation on the D-brane configurations yields the L\"uscher term  and allows us to confirm a previously proposed universal expression following from holography.
\end{abstract}

\newpage

 \numberwithin{equation}{section}

\section{Introduction}
The AdS/CFT correspondence has provided a powerful window into the strong coupling dynamics of gauge theories by proposing an alternative description in terms of supergravity theories~\cite{Maldacena:1997re,Witten:1998qj,Gubser:1998bc,Aharony:1999ti}. A particularly hopeful enterprize has been the search for models with properties resembling those of Quantum Chromodynamics (QCD). Important supergravity models dual to confining gauge theories have been constructed and shown to produce interesting strong coupling properties.  The best known examples are dual to field theories in 3 and 4 dimensions and they include: the Klebanov and Strassler model (KS)~\cite{Klebanov:2000hb}, the Maldacena-N\'u\~nez's (MN) interpretation~\cite{Maldacena:2000yy} of Chamsedine-Volkov~\cite{Chamseddine:2001hk} background,  Maldacena-Nastase (MNa)~\cite{Maldacena:2001pb}, and Cveti$\check{c}$, Gibbons, L\"u, and Pope (CGLP)~\cite{Cvetic:2001ma}.

An important QCD configuration are $k$-strings:  colorless combinations of quark-antiquark pairs stretched a distance $L$ which is much larger than the spatial separation $\varepsilon$ between the individual pairs. The energy of this configurations is proportional to $L$ and the coefficient of proportionality is the $k$-string tension (see \cite{Shifman:2005eb} for a review). In the holographic context the study of these configurations was initiated in \cite{Herzog:2001fq,Herzog:2002ss}; subsequent works include \cite{PandoZayas:2008hw,Doran:2009pp,Stiffler:2009ma,Stiffler:2010pz}.

Conformal field theories also have configurations analogous to $k$-strings in confining field theories. Important and clarifying aspects of these configurations were worked out for their analogues in the context of ${\cal N}=4$ Supersymmetric Yang-Mills/IIB string theory on $AdS_5\times S^5$ duality. In this context the $k$-string configurations are interpreted as Wilson loops in higher representations. For example, \cite{Drukker:2005kx}  proposed that the best description of Wilson loops in higher representations is achieved, on the dual gravity side, by D-branes with electric flux on their worldvolumes. A solid proof of the identification of Wilson loops in higher representations with D-branes with flux in their worldvolume was provided in \cite{Gomis:2006sb,Gomis:2006im} who concluded that Wilson loops in the fundamental representation are best described by a fundamental string, while the symmetric representation is described  by a $D3$-brane, and the anti-symmetric by a $D5$-brane. More general representations are, in principle, described by a set of D3 branes or  a set of D5 branes. Recently, the one-loop effective actions of Wilson loops in higher representations have been investigated in the holographic context of $D$-branes with fluxes in $AdS_5 \times S^5$ ~\cite{Faraggi:2011bb,Faraggi:2011ge}.

One of our goals is to extend the rigorous results of the conformal case to the realm of confining theories and to ultimately  connect our results with those of other approaches.

In \cite{Doran:2009pp}, we compared $k$-strings in two different gauge/gravity dual theories, one of $D4$-branes in the CGLP background and the other of $D3$-branes in the MNa background. In this work we find that a $D5$ brane embedded in either the MNa or the MN background has a solution whose tension exhibits $k$-ality and approximates a \emph{Casimir} law. This result is in stark contrast with the case of D3-branes on these backgrounds which yield exact \emph{sine} laws. It is also interesting that the MN is dual to a $4D$ $k$-string and the MNa is dual to a $3D$ $k$-string, and that they both yield the exact same tensions. In this paper we continue our program of holographic studies of $k$-strings by providing a unified treatment of D5-branes with worldvolume flux in two supergravity backgrounds MNa~\cite{Maldacena:2000yy} and MN~\cite{Maldacena:2001pb}.

One of the lessons we learned from the conformal case \cite{Gomis:2006sb,Gomis:2006im} is that D5 probe branes in a D5 generated supergravity background describe objects in the totally symmetric representation. When applied to our case, this fact is reflected in the D5 probe yielding a new tension law with values higher than both the Casimir and the \emph{sine} laws.  This interpretation is confirmed in the context of $k$-strings in $2+1$ theories where we can compare with data from lattice gauge theories and Hamiltonian methods.

 Finally, the quantum treatment of the branes yields the quantum correction to the $k$-string energy known as the L\"uscher term, which fits into a general formula for all other brane configurations we have computed. This provides further evidence for the universality class that these supergravity configurations form for $k$-strings.

This paper is organized as follows. In section~\ref{sec:D5Classical} we present the classical solution of a  $D5$ brane with electric flux embedded in the MN and MNa backgrounds. We find the exact same tension law for both embeddings. The formula  looks similar to a Casimir law but it is more complicated. In section~\ref{sec:CompTensions} we compare this new tension result to other brane embeddings that have been investigated~\cite{Herzog:2001fq,Herzog:2002ss} as well as lattice~\cite{Lucini:2001nv,Lucini:2002wg,Bringoltz:2006gp} and Hamiltonian results~
\cite{Karabali:2007mr}. In section~\ref{sec:D5Quantum} we show that the quantum D5 brane analysis agrees  with all of our previous analysis for the L\"uscher term  and they  all fit in an encompassing formula~(\ref{eq:LuscherGG}). Section~\ref{sec:conclusion} contains our conclusions. We relegate various technical aspects of the quadratic fluctuations around the classical solutions to a series of appendices.

\section{The Classical D5-brane in MN/MNa Backgrounds}\label{sec:D5Classical}


\subsection{The D$\it{p}$-brane action}
The 10-D bosonic backgrounds consists of a metric which is sourced by a Neveu-Scharz form, Ramond-Ramond forms, the dilaton and classically no fermion contributions:
\begin{align}
ds^{2}_{10}=G_{\mu \nu}dX^{\mu}dX^{\nu}, \qquad  H_{3}=dB_{2}, \qquad  F_{n+1}=dC_{n},\  \qquad \Phi.
 \end{align}
We embed a probe D$\it{p}$-brane at bosonic coordinates $X^{\mu}=X^{\mu}(\xi^{a})$ with  world volume coordinates $\xi^{a}$. We denote the brane's $U(1)$ gauge field as
\begin{align}
\mathcal{F}_{ab} = B_{ab}+2\pi {\alpha^{\prime}} F_{ab},
\end{align}
here $B_{ab}$ is the pullback of $B_{\mu\nu}$
\begin{align}
B_{ab} = B_{\mu\nu}\frac{\partial X^{\mu}}{\partial \xi^{a}}\frac{\partial X^{\nu}}{\partial \xi^{b}},
\end{align}
and $F_{ab}$ is a $U(1)$ gauge field on the brane
\begin{align}
F_{ab} = \frac{\partial A_{a}}{\partial \xi^{b}}-\frac{\partial A_{b}}{\partial \xi^{a}}.
\end{align}
The D$\it{p}$-brane action, for the approximation we consider is composed of  a Born-Infeld (BI) and a  Chern-Simons (CS) terms:
\begin{align}
S_{p} = S_{p}^{(BI)} + S_{p}^{(CS)} ~~~.
\end{align}
The BI piece of the action is
\begin{equation}
S_{p}^{(BI)} = -\mu_{p}\int{d^{p+1}\xi e^{-\Phi}\sqrt{-\text{det}\mathcal{M}_{ab}}}
\end{equation}
where
\begin{align}
\mathcal{M}_{ab} \equiv g_{ab} + \mathcal{F}_{ab}, \quad  \mathcal{M} \equiv -\text{det}(\mathcal{M}_{ab}),\quad
\mu_p = (2 \pi)^{-p} (\alpha')^{-(p+1)/2}
\end{align}
and $g_{ab} = g_{ab}(X(\xi))$ is the pullback of the 10-D metric $G_{\mu\nu} = G_{\mu\nu}(X(\xi))$
\begin{align}
g_{ab} = G_{\mu\nu}\frac{\partial X^{\mu}}{\partial \xi^{a}}\frac{\partial X^{\nu}}{\partial \xi^{b}}~~~.
\end{align}
The Chern-Simons action is written as
\begin{align}
S_{p}^{(CS)} &= \mu_{p}\int{}\sum_{{n}}e^{\mathcal{F}}\wedge C_{n} \notag \\  &= \mu_{p}\int{\left(C_{p+1}+\mathcal{F}\wedge C_{p-1}+\frac{1}{2}\mathcal{F}\wedge\mathcal{F}\wedge C_{p-3}+\ldots\right),}
\end{align}
where the $C_{n}$ are understood as the pullbacks of the Ramond-Ramond forms,
\begin{align}
C_{a_{1}\ldots a_{n}} = C_{\mu_{1}\ldots \mu_{n}} \frac{\partial X^{\mu_{1}}}{\partial \xi^{a_{1}}} \cdots \frac{\partial X^{\mu_{n}}}{\partial \xi^{a_{n}}}.
\end{align}
The sum over $\it{n}$ is dependent on the specific Sugra theory. Type IIB has $\it{n} = 0,2,4$ and type IIA has $\it{n} = 1,3$. Our action defines a classical field theory of $\it{scalar}$ fields, $X^{\mu} = X^{\mu}(\xi)$ and $U(1)$ $\it{gauge}$ fields, $A_{a} = A_{a}(\xi)$ on the D$\it{p}$-brane. This is, in fact, precisely one of the two parameters that we use to describe our L\"uscher term formula~(\ref{eq:LuscherGG}). Namely, $p$, which is the spatial dimension of the D$p$-brane world volume. The other parameters we used is $d$, which is the space-time dimension of the effective dual field theory.

\subsection{The MN/MNa backgrounds and Source Forms}

It is well known from the works of \cite{Kinar:1998vq,Brandhuber:1998er}, that the point $\rho =0$ corresponds to the confining region in the dual gauge theory.  The effective fundamental tension is given by $T=g_{00}(r_{min})/(2\pi \alpha')$.

The original discussion of $k$-strings in $3+1$ and $2+1$ dimensions used this crucial fact to simplify the construction of holographically dual configurations~\cite{Herzog:2001fq,Herzog:2002ss}. Namely, the D-brane configurations were localized precisely at this confining point in the bulk.  Exploiting this insight, we construct solutions that are localized in the confining region. Since we also discuss the one-loop properties of the D-brane configurations in section~\ref{sec:D5Quantum}, we present the MN/MNa solutions including up to second order in an expansion  of $\rho$ about $\rho=0$, the confining point.

The backgrounds take the form

\begin{align}
ds^{2} &= e^{\Phi}\left[dx_{d}^{2} + N\alpha^{\prime}\Big[ d\rho^{2}+R^{2}d\Omega_{6-d}^{2}+\frac{1}{4}\left(\omega^{a}-A^{a}\right)^{2}\Big]\right],\\
e^{\Phi} &= e^{\Phi_{0}}(1+c_{1}\rho^{2})+\mathcal{O}(\rho^{3}),\qquad R^{2} = \rho{^2}+\mathcal{O}(\rho^{3})\label{Rlimit},\\
F_{3} &= dC_{2}
= -\frac{N}{4}(\omega^{1}-A^{1})\wedge(\omega^{2}-A^{2})\wedge(\omega^{3}-A^{3})+\notag \\
&\qquad+\frac{N}{4}\sum_{{a}}F^{a}\wedge(\omega^{a}-A^{a})^{2}+c_{2}\frac{N}{16}\mathcal{O}(\rho^{4}),
\end{align}
where
\begin{align}
\sigma^{a}A^{a} &= \sigma^{1}a(\rho)d\Theta_1+\sigma^{2}a(\rho)\text{sin}\Theta_1 d\Phi_1+\sigma^{3}b(\rho)\text{cos(}\Theta_1 )d\Phi_1,\notag \\
\omega^{1}+i\omega^{2} &= e^{2i\Psi_2}(d\Theta_2+\text{sin}\Theta_2 d\Phi_2),\qquad
\omega^{3} = -2d\Psi_2+\text{cos(}\Theta_2 )d\Phi_2 ,\notag \\
a(\rho) &= 1+a_{2}\rho^{2}+\mathcal{O}(\rho^{3}),\qquad b(\rho) = 1+b_{2}\rho^{2}+\mathcal{O}(\rho^{3}),
\end{align}
where $\sigma^a$ are the Pauli matrices. The values of the parameters for the MN background fields defined above are given by
\begin{align}
d = 4,\quad c_{1} = \frac{4}{9},\quad c_{2} = 0,\quad a_{2} = -\frac{2}{3}, \quad b_{2} = 0 \notag
\end{align}
and the MNa parameters are
\begin{align}
d = 3,\quad c_{1} = \frac{7}{24},\quad c_{2} = 1,\quad a_{2}= b_2 = -\frac{1}{6},  \notag
\end{align}
We also choose the gauge for the R-R two form as
\begin{align}
C_{2} = \frac{N\alpha^{\prime}}{4} \Big[ a(\rho)\text{cos(}2\Psi_2 )d\Theta_1 \wedge d\Theta_2 + 2\Psi_2 b(\rho)\text{sin(}\Theta_1 )d\Theta_1 \wedge d\Phi_1 -\notag \\ a(\rho)\text{sin(}\Theta_2) \text{sin(}2\Psi_2 )d\Theta_1 \wedge d\Phi_2  -a(\rho)\text{sin}\Theta_1 \text{sin(}2\Psi_2 )d\Theta_2 \wedge d\Phi_1 + \notag \\
2\Psi_2 \text{sin(}\Theta_2 )d\Theta_2 \wedge d\Phi_2 +2b^{\prime}(\rho)\Psi_2 \text{cos(}\Theta_1 )d\Phi_1 \wedge d\rho+\notag \\
\Big( b(\rho)\text{cos(}\Theta_1  \text{)cos(}\Theta_2)+a(\rho)\text{cos(}2\Psi_2 \text{)sin}\Theta_1 \text{sin}\Theta_2 \Big) d\Phi_1 \wedge d\Phi_2 \Big].
\end{align}

\subsection{The $k$-string Tension Law}

\paragraph{} Let us briefly summarize our results and place them in the bigger frame of the $k$-string literature.  We now embed a D5-brane probe in the MN and MNa type IIB SUGRA backgrounds, and extract the tension from its classical energy. The embedding is different for each background but we show that the Hamiltonians, and thus the string tensions, are identical. The solution corresponds to a nontrivial embedding and its worldvolume topology is  $ {\mathbb R}^{1,1} \times \mathcal{I} \times S^3$, where $\mathcal{I}$ is an interval of ${\mathbb R}^1$. This topology contrasts with those investigated in~\cite{Herzog:2001fq} and~\cite{Herzog:2002ss} which were ${\mathbb R}^{1,1} \times S^2$ and ${\mathbb R}^{1,1} \times S^3$, respectively. We integrate out the angular degrees of freedom to obtain an effective string. For the purpose of string tensions, we pass to the Hamiltonian formalism via Legendre transformation. Solving for the conjugate momentum in terms of a constant electric flux on the brane, substituting the expression back into the Hamiltonian and then extremizing with respect to a background parameter present on the D5-brane leads to the brane tension which we interpret as the field theory $k$-string tension.

Recall that $k$-strings are open strings with their ends fixed on the boundary. Here, however, we focus on the portion of the string localized at $\rho=0$ because the part of the action coming from the extension to the boundary where the field theory lives is interpreted as describing the infinite mass of external quarks \cite{Kruczenski:2004me}, we thus regularized the tension by subtracting that piece.

The holographic configuration that best captures the properties of $k$-strings is constructed as follows. First, as a string in the dual field theory, we expect it to be extended along one spatial dimension, that is, to live in $(t,x)$. Further, as follows from previous analysis we want to include a $U(1)$ gauge field in the worldvolume of the brane that represents the number of fundamental strings dissolved in the worldvolume  \cite{Pawelczyk:2000hy} \cite{Camino:2001at}. The $k$-string tension is identified with the classical tension on the D-brane  configuration. As a consistency check we verify that the resulting tension satisfy the $k$-ality condition as dictated by the representation theory of the field theory configuration.

\subsubsection{Tension From the MN/MNa  backgrounds}
\indent The first step is to consider the pullback of the MN/MNa backgrounds to the D5-brane  worldvolume in  the limit in which  the holographic coordinate to zero, $\rho$, goes to zero.  We note that while the MN background is dual to 3+1 dimensional field theory and the MNa background is dual to a 2+1 dimensional one, our solutions produce the exact same tension law. Below is the MN calculation. The MNa background calculation is completely analogous to the MN case with only a slightly difference embedding map and metric. Referring to the metric, for the MN case assume a constant mapping solution of $\Psi_{2}$ on the brane, with explicit coordinate mappings below.
The D5-brane parameterization is initially given by
\begin{align}
\xi^{a} = (t,x,\theta_{1},\theta_{2},\phi_{1},\phi_{2})
\end{align}
As we will illustrate below, it is important to choose an embedding that guarantees that the worldvolume metric is non-degenerate.  This criterion requires that $X^2$ in the MNa case and $X^3$ in the MN case be suitable functions of the angles $\theta_{1},\theta_{2},\phi_{1},\mbox{and\,\,}\phi_{2}$  in the embeddings. For example, an embedding which has $X^3= X^3_0 $, were $X^3_0$ is a constant as in\begin{align}
X_{MN}^{\mu} &= (X^{0},X^{1},X^{2},X^3,\Theta_{1},\Theta_{2},\Phi_{1},\Phi_{2},\Psi_{2},\rho) \notag \\
&= (t,x,0,X^3_0 ,\theta_1,\theta_2,\phi_{1},\phi_{2},\psi_{2_0},0) \label{badchoice}
\end{align}
will yield a
metric whose determinate is proportional to $R^{2}$.  From Eq.(\ref{Rlimit}) one sees that as $\rho\rightarrow 0$, the determinate will vanish.

Examples of embeddings that induce finite volume D5 branes at $\rho=0$ are given below. It will be instructive to consider two cases:

\begin{enumerate}
\item \begin{align}
X_{MN}^{\mu} &= (X^{0},X^{1},X^{2},X^3,\Theta_{1},\Theta_{2},\Phi_{1},\Phi_{2},\Psi_{2},\rho) \notag \\
&= (t,x,0,X^3_0 (\phi_++\phi_-),\theta_1,\theta_2,\phi_{1},\phi_{2},\psi_{2_0},0)
\end{align}
and
\begin{align}
X_{MNa}^{\mu} &= (X^{0},X^{1},X^{2},\Psi_1,\Theta_{1},\Theta_{2},\Phi_{1},\Phi_{2},\Psi_{2},\rho) \notag \\
&= (t,x,X^2_0(\phi_++\phi_-),0,\theta_1,\theta_2,\phi_{1},\phi_{2},\psi_{2_0},0),
\end{align}
which gives a metric with finite volume at $\rho=0$ of
$$
\text{Vol}^{1}_{D5}=4\pi^2 (N \alpha')^{\frac{3}{2}}\exp{(3 \Phi_0)}X^i_0 \sin{(2 \psi_{2_0})}
$$
for both the MN and MNa cases.   Here $i=2,3$ for the MNa and MN cases respectively.
\item The case we will explore throughout this work uses the embedding,
\begin{align}
X_{MN}^{\mu} &= (X^{0},X^{1},X^{2},X^3,\Theta_{1},\Theta_{2},\Phi_{1},\Phi_{2},\Psi_{2},\rho) \notag \\
&= (t,x,0,-X^3_0 \cos(\theta_+)\cos(\theta_-),\theta_+,\theta_-,\phi_{1},\phi_{2},\psi_{2_0},0)
\end{align}
and
\begin{align}
X_{MNa}^{\mu} &= (X^{0},X^{1},X^{2},\Psi_1,\Theta_{1},\Theta_{2},\Phi_{1},\Phi_{2},\Psi_{2},\rho) \notag \\
&= (t,x,-X^2_0\cos(\theta_+)\cos(\theta_-),0,\theta_+,\theta_-,\phi_{1},\phi_{2},\psi_{2_0},0),
\end{align}
where above we have denoted
$$
\theta_+= \frac{\theta_1+\theta_2}{2},\,\,\,\,\,\,\,\,\, \theta_-= \frac{\theta_1-\theta_2}{2}.
$$
For this case, the induced worldvolume is
$$
\text{Vol}^{2}_{D5}=\pi^2 (N \alpha')^{\frac{3}{2}}\exp{(3 \Phi_0)}X^i_0 \cos{(2 \psi_{2_0})}
$$
for the same interpretation of $X^i$.
In this case the nontrivial embedding of the brane in the $X^2$ coordinate for MNa and $X^3$ for MN is a mapping  into a segment on the D5 brane.
\end{enumerate}

In both cases the constant $\psi_{2_0}$ is an extremized value for $\Psi_2$. The values that minimize  the Hamiltonian via,
\begin{equation}\label{eq:Hconstraint}
\partial_{\Psi_{2}}\mathcal{H}(\Psi_{2})\vert_{\psi_{2_0}} = 0
\end{equation}
 are $\psi_{2_0}^{\text{min}}=\pi (n+\frac{1}{4})$ for Case 1 above and  $\psi_{2_0}^{\text{min}}=\pi (n+\frac{1}{2})$ for Case 2.

Interestingly enough, both of these examples  will give the same tension laws.  Let us restrict our attention to \uline{Case 2} from here on out.  Then the pullback of the metric on the D5-brane for both MN and MNa backgrounds is:
\begin{align}
\begin{split}
ds_{MN}^2=& \frac{1}{8} e^{\Phi_0} \bigg(2 N \alpha ' \left(d\theta_{1}^2+d\phi_{1}^2-2 d\phi_{1} d\phi{2} \cos (\theta_{1})+d\phi_{2}^2 \right)\\
-&8 dt^2 + 8 dx^2  + 2(X_0^3)^2\left(\sin\theta_1 d\theta_1 +\sin\theta_2 d \theta_2\right)^2\bigg) \\
\end{split}
\end{align}
and
\begin{align}
\begin{split}
ds_{MNa}^2=& \frac{1}{8} e^{\Phi_0} \bigg(2 N \alpha ' \left(d\theta_{1}^2+d\phi_{1}^2-2 d\phi_{1} d\phi{2} \cos (\theta_{1})+d\phi_{2}^2 \right)\\
-&8 dt^2 + 8 dx^2  + 2(X_0^2)^2\left(\sin\theta_1 d\theta_1 +\sin\theta_2 d \theta_2\right)^2\bigg) \\
\end{split}
\end{align}
Both worldvolume induced metrics have world volumes forms given by
\begin{equation}
dV = r_{\text{effective}}^4 \sin{(\theta_1)}\sin{(\theta_2)} dx\,dt\,d\theta_1\,d\theta_2\,d\phi_1\,d\phi_2,
\end{equation}
 where the effective radii, $r_{\text{effective}}$,  for the respective backgrounds are given by
\begin{equation}
 r^4_{\text MN} =\frac{1}{ 16}{(N \alpha ')}^{3/2} e^{3 \Phi_0}{X^3_0} \cos({2{\psi_2}_0}), \quad
 r^4_{\text MNa} =\frac{1}{ 16}{(N \alpha ')}^{3/2} e^{3 \Phi_0}{X^2_0} \cos({2{\psi_2}_0}) . \label{volumes}
\end{equation}

Both the MN and MNa have induced geometries with  a scalar curvature given by
\begin{equation}
R= \frac{6}{N \alpha '} e^{-\Phi_0}, \label{scalar}
\end{equation}
which is small in the limit of parameters that we choose for the solution and attests to the validity of the supergravity background  and Born-Infeld  action in the given approximations.
From the above we calculate the D5-brane action below. Consider the action,
\begin{align}
S_{5}^{(full)} = -\mu_{5}\int{d^{6}\xi e^{-\Phi}\sqrt{-\text{det}\mathcal{M}_{ab}}} + \mu_{5}\int{\left(\frac{1}{2}\mathcal{F}\wedge\mathcal{F}\wedge C_{2}\right)}
\end{align}
where,
\begin{align}
\mathcal{M}_{ab} \equiv g_{ab} + \mathcal{F}_{ab}
\end{align}
with the only non-vanishing components of $\mathcal{F}_{ab}$ being $\mathcal{F}_{10} = - \mathcal{F}_{01} = 2 \pi \alpha' E$.

In the above $E$ is the electric field. The D5-brane Lagrangian density reduces to
\begin{equation}
\mathcal{L}=-\frac{N^{3/2} e^{\Phi_{0}}(X^i_0) \sin (\theta_{1}) \sin (\theta_{2}) \cos (2 \psi
   _{2_0}) \sqrt{e^{2 \Phi_{0}}-4 \pi ^2 \text{E}^2 \alpha '^2}}{512 \pi ^5 \alpha '^{3/2}}
\end{equation}
where the subindex $i$ takes values $i=2$ for MNa and $i=3$ for MN. For our purposes we integrate out the angular degrees of freedom. This choice is available to us due to our judicious choice electric field and dilaton evaluated at the holographic coordinate $\rho =0$. We next perform a Legendre transformation in order to obtain the D5-brane Hamiltonian
\begin{align}
\mathcal{H} = E\frac{\partial \mathcal{L}}{\partial\dot{A}} - \mathcal{L}.
\end{align}
We equate the conjugate momentum to a constant $\Pi$ by definition such that $\frac{\partial \mathcal{L}}{\partial\dot{A}} \equiv \Pi $, with $\dot{A} = E$. We interpret this transformation along the lines of \cite{Craps:1999nc,Camino:1999xx} and find,
\begin{equation}
\Pi =\frac{\text{E} N^{3/2} e^{\Phi_{0}} X^i_0 \sqrt{\alpha '} \cos (2 \psi_{2_0})}{8 \pi  \sqrt{e^{2\Phi_{0}}-4 \pi ^2 \text{E}^2 \alpha '^2}}.
\end{equation}
Since $E$ has mass dimensions $[E]= Mass^2$ and $\alpha'$ has dimensions $[\alpha']= 1/Mass^2$, and $[X^i_0]=1/Mass$, we see that  $\Pi$ is dimensionless.
Using the conjugate momentum defined in the previous line, we solve for the electric field in terms of the conjugate momentum which has both positive and negative roots. Note that there exists a distinct Hamiltonian for each root.  We solve for each root solution for the electric field and  substitute back the solution in the Hamiltonian. Thus we have
\begin{equation}\label{eq:Hminus1}
\mathcal{H}_{-} = \frac{e^{\Phi_{0}} \left(N^3 e^{2 \Phi_{0}} (X^i_0)^2 \cos ^2(2 \psi_{2_0})-256 \pi ^4 \Pi^2 \alpha '\right)}{32 \pi ^3 \alpha '^{3/2} \sqrt{256 \pi ^4 \Pi^2 \alpha '+N^3 e^{2 \Phi_{0}} (X^i_0)^2 \cos ^2(2 \psi_{2_0})}}
\end{equation}
While for the positive electric field solution we have,
\begin{equation}\label{eq:Hplus1}
\mathcal{H}_{+} = \frac{e^{\Phi_{0}} \sqrt{256 \pi ^4 \Pi^2 \alpha '+N^3 e^{2 \Phi_{0}} (X^i_0)^2 \cos ^2(2
   \psi_{2_0})}}{32 \pi ^3 \alpha '^{3/2}}
\end{equation}

It is clear that $\mathcal{H}_{+}$ has a minimum only when $\cos{(2 \psi_{2_0})}$ vanishes. But as one can see in Eq.({\ref{volumes}}), such a minimum is singular as the value of the metric determinant on the  D5 worldvolume vanishes.
  Therefore we focus on the
 $\mathcal{H}_{-}$ solution where one can show that this solution exhibits $k$-ality and  can be identified with fundamental charge dissolved on the 5-brane worldvolume  \cite{Craps:1999nc,Camino:1999xx}. We wish to minimize $\mathcal{H}_{-}$ with respect to $\psi_{2_0}$ in order to solve for the string tension in terms of an extremized  value of $\psi_{2_0}$ consistent with Eq.(\ref{eq:Hconstraint}). This yields a family of solutions of the form
\begin{equation}\label{eq:psi20}
\psi_{2_0} = \frac{\pi}{2} n  \qquad \text{for} \quad n\in \mathbb{Z}.
\end{equation}
Note that taking $n\to n+1$ leaves the volume form invariant but has the effect of exchanging $\theta_1 \leftrightarrow \theta_2$.
\subsubsection{ The Tension Law}\label{sec:TenLaw}

Inserting the solution in Eq.~(\ref{eq:psi20}) back into the Hamiltonian, Eq.~(\ref{eq:Hminus1}) yields the tension on the D5-brane:
\begin{equation}\label{eq:Hmin}
T^i = \frac{e^{\Phi_{0}} \left(N^3 e^{2 \Phi_{0}} (X^i_0)^2-256 \pi ^4 \Pi^2 \alpha '\right)}{32 \pi ^3
   \alpha^{3/2} \sqrt{256 \pi ^4 \Pi^2 \alpha '+N^3 e^{2 \Phi_{0}} (X^i_0)^2}}
\end{equation}
with the label $i = 2$ for MNa and $i=3$ for MN.
We see that the form of the 5-brane tension is the same for either background, the only difference being the value of the bulk fields $X_0^i$ on the brane.

\begin{figure}[htp]
\begin{center}
\includegraphics[width=0.5\textwidth]{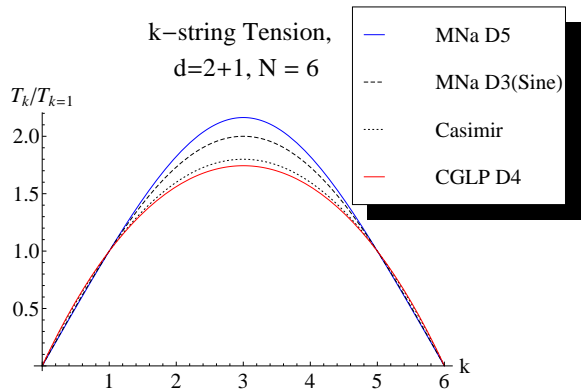}
\caption{\label{fig:tension2p1N6}The $N=6$, $d=2+1$ $k$-string tension for various gauge/gravity models compared to the Casimir and sine laws. Clearly, the MNa $D5$ representation is the highest energy representation of all of these models, just as in the $N=4$ case.  }
\end{center}
\end{figure}

\begin{figure}[htp]
\begin{center}
\includegraphics[width=0.5\textwidth]{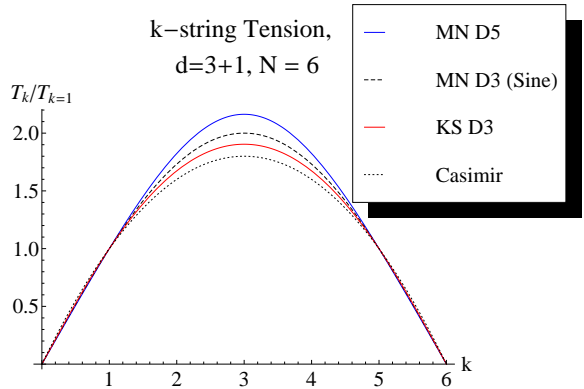}
\caption{\label{fig:tension3p1N6}The $N=6$, $d=3+1$ $k$-string tension for various gauge/gravity models compared to the Casimir and sine laws. Clearly, the MNa $D5$ representation is the highest energy representation of all of these models. Comparing with recent lattice data~\cite{Athenodorou:2010cs}, we see that the MNa D5-brane is acting more like the symmetric representation, where as the KS D3-brane is acting more like the symmetric representation. The MN D3-brane, which is a precise sine law, is in between.}
\end{center}
\end{figure}

Since the Euler-Lagrange equations require the  conjugate momentum to be a constant and anticipating a solution that enjoys $N$-ality, we write the quantization condition $\Pi = k$.   Then with this, the correct choice in the pullback parameter,$ X^i_0$, that ensures $N$-ality is $X^i_0= 16 \pi^2 e^{-\Phi_0} \sqrt{\frac{\alpha'}{N}}$.  From Eq. (\ref{eq:Hmin}) one can see that this choice for $X^i_0$ will give zero tension when $k=N$.
Upon substitution of these values back into their respective tension laws  we obtain the exact same D5-brane tensions for both MN and MNa backgrounds
\begin{align}
T_{MN/MNa} = \frac{e^{\Phi_{0}} (N-k) (k+N)}{2 \pi  \alpha ' \sqrt{k^2+N^2}}.
\end{align}
We rescale $k\to 2k_{0}-N$ which allows us to come to the final D5-brane tension,
\begin{align}
T_{D5} = \frac{\sqrt{2} e^{\Phi_{0}} k_{0}(N-k_0)}{\pi  \alpha' \sqrt{k_0^2 + (N-k_0)^2}}. \label{tension}
\end{align}
Figures~\ref{fig:tension2p1N6} and \ref{fig:tension3p1N6}  show that this tension exhibits $N$-ality, and is larger than both the Casimir and Sine laws. Comparing to the data from Table~\ref{tab:comparetensions}, we see that this aligns qualitatively with the identification as $D5$-branes being the symmetric representation and $D3$-branes being the anti-symmetric representation in the case of MNa.

In what follows we present holographic results for the $k$-string tension that follow from the present work and all from our previous investigations \cite{PandoZayas:2008hw,Doran:2009pp,Stiffler:2009ma,Stiffler:2010pz}. In all we present the $k$-string tensions in 2+1 and 3+1 dimensional field theories using probe D3, D4 and D5 branes.

In Fig. \ref{fig:tension2p1N6}, we present the holographic $k$-string tension computed for 2+1 field theories using the a D5-brane in the MNa solutions (this work), D3 in the MNa solutions, and a D4 in the CGLP solutions which was computed \cite{Herzog:2002ss}; we have also plotted the Casimir law to orient the reader.

In Fig. \ref{fig:tension3p1N6} we consider the results for 3+1 dimensional theories. We have included the results of D5 brane in the MN background (this work), a D3 brane in the MN solution yielding precisely a {\it sine} law; we also consider the D3 probe brane in the Klebanov-Strassler background presented in \cite{Herzog:2001fq}. Finally, we have also plotted the Casimir law to guide the reader.

In the next section we put our results in the context of higher representations and compare them to results provided by other methods, when available.

\section{Tensions from Various Methods}\label{sec:CompTensions}

In this section we compare the results of $k$-string tensions from holographic computations with those obtained using various other approaches.

Computations of $k$-string tensions have been performed in various frameworks.  In the context of lattice gauge theories we quote the most recent results due to Bringoltz and Teper \cite{Bringoltz:2006gp}. For other methods it is quite challenging to address the question of $k$-string tensions. In a very interesting work, \cite{Douglas:1995nw}, Douglas and Shenker found a {\it sine law} for $k$-strings in the context of Seiberg-Witten theories (see also \cite{Armoni:2003ji} for a more comprehensive discussion, and \cite{Shifman:2005eb} for a general review). General results for confining QCD-like theories are in general lacking. One beautiful exception is the work of Karabali and Nair who used a Hamiltonian approach to compute $k$-string tensions in $2+1$-dimensional Yang-Mills \cite{Karabali:2007mr}. Their full answer states the $k$-string tension follows precisely a {\it Casimir law}. Couplings of Yang-Mills to matter in this framework has been also presented in \cite{Agarwal:2007ns,Agarwal:2009gb,Agarwal:2012bn}. An interesting work using different methods but extending the 3d YM calculation to 3d YM with adjoint matter was recently presented by Armoni-Dorigoni-Veneziano \cite{Armoni:2011dw}. The paper uses the Eguchi-Kawai volume reduction to calculate the tension of $k$-strings in the theories with adjoint fermions and obtains a sine law, $T_k=N\sin(\pi \, k/N)$.

Following the discussion of Gomis and Passerini \cite{Gomis:2006sb,Gomis:2006im}, we identified a probe D5 in the Maldacena limit of D3 background as configurations in the antisymmetric representation. Similarly, a D3 brane in the Maldacena limit of a D3 brane background or a D5 brane in the Maldacena limit of a D5 background corresponds to the symmetric representation. This very general conclusion is based on the analysis of $Dp/Dq$ brane bound states discussed in Polchinski's string theory monography \cite{Polchinski:1998rr} and was explicitly spelled out in \cite{Gomis:2006sb,Gomis:2006im}.

\begin{table}[htp]
\centering
\begin{tabular}{|c|c|c|c|c|c|}
\multicolumn{6}{c}{$T_k/T_1$ from Various Methods in $2+1$} \\
\multicolumn{6}{c}{S=symmetric, A=antisymmetric, M=mixed} \\
\hline
Group & $k$ &  CGLP~\cite{Herzog:2002ss} & MNa  & BT\cite{Lucini:2001nv,Lucini:2002wg,Bringoltz:2006gp} & KN~\cite{Karabali:2007mr}  \\
\hline\hline
\multirow{2}{*}{$SU(4)$} & \multirow{2}{*}{2}  & 1.310(A) & 1.414 (A)  &  1.353(A) & 1.333(A)   \\
&&& 1.491 (S) & 2.139(S) & 2.400(S)\\
\hline\hline
\multirow{2}{*}{$SU(5)$} & \multirow{2}{*}{2} &  1.466(A) & 1.618 (A) & 1.529(A) & 1.5 (A)  \\
&&& 1.715(S) &&\\
\hline\hline
\multirow{5}{*}{$SU(6)$} & \multirow{2}{*}{2} & 1.562(A) & 1.732(A) & 1.617(A) & 1.6(A) \\
&&& 1.824(S) & 2.190(S) & 2.286(S)\\
\cline{2-6}
& \multirow{3}{*}{3} & 1.744(A) & 2.0(A)  & 1.808(A) & 1.800(A) \\
&&& 2.163(S) &3.721(S) & 3.859(S) \\
&&&&2.710(M) & 2.830(M)\\
\hline\hline
\multirow{6}{*}{$SU(8)$} & \multirow{2}{*}{2} & {1.674(A)} & 1.848 (A)  & 1.752(A) & 1.714(A)\\
&&& 1.917(S) &&\\
\cline{2-6}
&\multirow{2}{*}{3} & 2.060(A) & 2.414 (A) & 2.174(A) & 2.143(A)\\
&&& 2.599(S) &&\\
\cline{2-6}
&\multirow{2}{*}{4} & 2.194(A) & 2.613(A) & 2.366(A) & 2.286(A)\\
&&& 2.857(S) &&\\
\hline
\end{tabular}
\caption{Comparison of $2+1$ $k$-string tensions from various methods.  The values quoted are $T_k/T_1$, where $T_k$ is the $k$-string tension, and $T_1 = T_{k=1}$ is the $k=1$ string tension. }
\label{tab:comparetensions}
\end{table}

Let us conclude this question by addressing an important question\footnote{We are indebted to Adi Armoni for raising this question and the interesting discussion that ensued.}. It is believed that in a confining theory the tension of the $k$-symmetric string and the $k$-antisymmetric strings are the same. Screening turns the symmetric source into an antisymmetric. This view is defended, for example in \cite{Armoni.Talk:2010}. What transpire from figures \ref{fig:tension2p1N6} and \ref{fig:tension3p1N6}, is that the $k$-symmetric representation described holographically by  MN D5 and MNa D5 has higher tension. Presumably these D5 configurations, having higher energy, will convert themselves dynamically into D3 configurations.

Such classical solutions, if they exist, are rather complicated and will require methods beyond the scope of this paper.  For example, the solutions should be time-dependent and interpolate between one brane as $t\to -\infty$ and another as $t\to \infty$; note that the boundary conditions involved are different in dimensionality. The situation then suggests that it is logically possible that the holographic configurations  described in this manuscript that correspond to $k$-symmetric strings in the dual field theory are metastable.

It is worth restating that the holographic calculation is valid at large $N$, namely, in the limit with $N\to \infty$, with $\lambda= g_{YM}^2 N$ fixed. Our intuition of screening can be very different in this limit. For example, the adjoint string discussed  \cite{Armoni.Talk:2010} can not break in this limit. If we borrow some intuition from the AdS/CFT correspondence in the case of ${\cal N}=4$ supersymmetric Yang-Mills where the corresponding configurations are Wilson loops in the appropriate representations with $N\to \infty$ and $k/N$ fixed, we conclude that each configuration is, at least, metastable in the 't Hooft limit.

\section{The Quantum D5-brane in MN/MNa Backgrounds}\label{sec:D5Quantum}

In this section we discuss aspects of the quantum fluctuations for the classical D-brane configurations corresponding to $k$-strings in the dual field theories.

\subsection{The Geometry of the Minimized Solution}\label{sec:Geometry}
The aforementioned values of $\psi_{2_0}$, that is,  $\psi_{2_0} = \frac{\pi}{2} n $, allows us to recast the minimized D5 brane metric into an ${\mathbb R}^{1,1}\times \mathcal{I} \times S^3$ geometry, where $\mathcal{I}$ is an interval manifold. The coordinates $(x,t,y)$ chart the  $\mathcal{M}^3={\mathbb R}^{1,1}\times \mathcal{I}$, while the $S^3$ is charted with Hopf coordinates $(\theta, \phi_+, \phi_-)$.  Consider the coordinate transformation on the D5  given by:
\begin{equation}
\xi^a=(t, x, \theta_1, \theta_2, \phi_1, \phi_2)\rightarrow\   \xi'^{a} = (t,x,y,\theta_{},\phi_{+},\phi_{_{-}}), \label{newcoordinates}
\end{equation}
where
\begin{eqnarray}
\theta_2 &\rightarrow& - \arccos{(2 y -\cos{(2 \theta)})},\nonumber\\
\theta_1& \rightarrow& 2 \,\theta,\\\nonumber
\phi_1 &\rightarrow& \phi_+-\phi_-,\\\nonumber
\phi_2 &\rightarrow&  \phi_++\phi_-\,.
\end{eqnarray}
The domain of the new coordinates is then, $(-1\le y\le 1), (0 \le \theta \le \frac{\pi}{2}), \, (0 \le \phi_+ \le \pi), $ and $ (\frac{\pi}{2} \le \phi_- \le \frac{\pi}{2})$.
These are the Hopf coordinates on $S^3$, modulo the phase shift of  $\phi_- \rightarrow \phi_-+ \frac{\pi}{2}$.  Then the metric  can be easily seen to have $\mathcal{M}^3 \times S^3$ geometry,
\begin{equation}
ds_{}^2 =e^{\Phi_0}\left( dx^2- dt^2 + ({X^j}_0 dy)^2\right) + 2 N \alpha 'e^{\Phi_0}  \left(d\theta_{}^2+\sin^2{\theta}\, d\phi_{+}^2+\cos^2{\theta} \,d\phi_{-}^2\right).
\end{equation}
Here
${X^j}_0 =8 \frac{\sqrt{\alpha'}}{N} \pi^2 \exp{(-\Phi_0)}$, for both the MN and MNa cases. The minimal D5 volume form  is then
\begin{equation}
dV_{\text min} = (2 N \alpha')^{3/2} \exp{(3 \Phi_0)} X^j_0 \cos{\theta} \sin{\theta} dx~dy~dt~d\theta~d\phi_+ ~ d\phi_-.
\end{equation}
  The $k$-string tension, Eq.[\ref{tension}], and scalar curvature, Eq.[\ref{scalar}], remain the same. This simplification of the metric will be useful for  finding explicit solutions to the perturbations as seen in the appendix.

\subsection{Quadratic fluctuations}

The stability of the configuration as well as features such as the L\"uscher term require that we examine the quadratic fluctuations of the $k$-string configuration about its classical solution.  Using the coordinates, $\xi' $ , defined  by Eq. (\ref{newcoordinates}), we fluctuate about the classical solution as
\begin{align}
X_{MN}^{\mu} &= (X^{0},X^{1},X^{2},X^3,\Theta_{1},\Theta_{2},\Phi_{1},\Phi_{2},\Psi_{2},\rho) \notag \\
&= \bigg(t,x,\lambda \delta X^2(\xi'),-X_0^3 \,y + \lambda \delta X^3(\xi'), \cr
&~~~~~\theta -\frac{1}{2} \arccos({2 y -\cos{(2 \theta)}}) + \lambda \delta\Theta_1(\xi'), \theta +\frac{1}{2} \arccos({2 y -\cos{(2 \theta)}})+ \lambda \delta \Theta_2(\xi'), \cr
&~~~~~\phi_{+} - \phi_-+ \lambda \delta \Phi_1(\xi'),\phi_{+} + \phi_-+ \lambda \delta \Phi_2(\xi')_,\psi_{2_0} + \lambda \delta \Psi_2(\xi'),\lambda \delta \rho(\xi') \bigg)
\end{align}
and
\begin{align}
X_{MNa}^{\mu} &= (X^{0},X^{1},X^{2},\Psi_1,\Theta_{1},\Theta_{2},\Phi_{1},\Phi_{2},\Psi_{2},\rho) \notag \\
&= \bigg(t,x,-X_0^2\,y+ \lambda \delta X^2(\xi),\lambda \delta \Psi_1(\xi'), \notag\\ &~~~~~ \theta -\frac{1}{2} \arccos({2 y -\cos{(2 \theta)}}) + \lambda \delta \Theta_1(\xi'),\theta +\frac{1}{2} \arccos({2 y -\cos{(2 \theta)}}) + \lambda \delta\Theta_2(\xi'),\notag\\
 &~~~~~\phi_{+} - \phi_-+\lambda \delta \Phi_+(\xi'),\phi_{+} + \phi_-+ \lambda \delta \Phi_-(\xi'),\psi_{2_0}+ \lambda \delta \Psi_2(\xi'),\lambda \delta \rho(\xi')\bigg)
\end{align}
with the gauge field fluctuations given by
\begin{align}
    A^\mu = (-\frac{1}{2} E\, x+ \lambda \delta A_t,\frac{1}{2} E\, t +  \lambda\delta A_x,  \lambda\delta A_{y},  \lambda\delta A_{\theta}, \lambda \delta A_{\phi_+},  \lambda\delta A_{\phi_-})
\end{align}
in both backgrounds, where $\lambda$ is an infinitesimal formal parameter to keep track of the order in perturbation theory.
 These fluctuations will produce an effective Lagrangian at order $\lambda^2$.   The first order in  $\lambda$   contribution should vanish, upon imposing the classical equations of motion, and up to total derivatives; any non-trivial contributions at this order represent further constraints on the perturbative fields.

\subsection{Effective Lagrangian }
In both, the MN and MNa,  cases the first order  Lagrangian  is a total derivative except for the  term
\begin{equation}
\mathcal{{L}}^{1st}_{\lambda} = 2 \cos(2\, \theta) \,\partial_{\theta}(\delta {A}_y). \label{straymagetism}
\end{equation}
This constrains $\delta A_{y}$ to either be a constant or to be symmetric in the interval $(0 \le \theta \le \pi) $, in order to avoid a  magnetic flux in the classical configuration. At second order in $\lambda$, the variations of  $\delta\Theta_1(\xi)$  and $\delta\Theta_2(\xi)$ for both cases only contribute to total derivatives.  We may use the diffeomorphism invariance to set the fluctuations , $\delta \Phi_+(\xi)$, $ \delta \Phi_-(\xi)$, and  $\delta \Psi_2(\xi)$ to zero.  If we choose to  include their fluctuations, these fields serve as Lagrange multiplier fields that demand that the magnetic fields, $F_{y\,\phi_\pm} = 0$. These constraints can be satisfied when   $\delta {A}_y$ vanishes.

With this, at  order $\lambda^2$, the fluctuations induce an effective metric on the D5 brane.   By choosing $\psi_{2_0}= \frac{\pi}{2}$ for the MN case and $\psi_{2_0}= 0$ for the MNa case, their effective metrics take the same form, in the coordinates $\xi' =(x,y,t, \theta, \phi_+,\phi_-)$,
\begin{align}
g_{a b}=
 \left(
\begin{array}{cccccc}
g_{11} & 0  & 0 & 0 & 0 & 0 \\
 0  & g_{22}   & 0 & 0 & 0 \\
 0 & 0 &-g_{11} & 0 & 0 & 0 \\
 0 & 0 & 0 & g_{23}  & 0   & 0 \\
 0 & 0 & 0 & 0    &g_{23} \sin{(\theta)}^2  & 0 \\
 0 & 0 & 0 & 0 & 0 & g_{23}\cos{(\theta)}^2 \\
\end{array}
\right)
\end{align}
with
\begin{equation}
g_{1 1}=\frac{e^{\Phi_0} N^2}{k^2+N^2}, \,\, g_{2 2}= \frac{1}{N} 64 e^{- \Phi_0} \pi^4  \alpha' , \,\, g_{2 3} =e^{\Phi_0}  N\alpha'.
 \end{equation}
 The differential volume form is $dV = 4 \pi^2 \sqrt{g_{11}} N {\alpha'}^2 \sin{(2 \theta)}dx dy dt d\theta d\phi_+ d\phi_-$.
Now we rescale the $(x,y,t)$ coordinates on    $ \mathcal{M}^3$ by writing   $(x\rightarrow \sqrt{g_{11}}\, x, \, y\rightarrow \sqrt{g_{22}}\, y, \, t\rightarrow \sqrt{g_{11}} \,t).   $ This rescaling allows us to write the metric  as,

\begin{align}
g_{a b}=
 \left(
\begin{array}{cccccc}
1 & 0  & 0 & 0 & 0 & 0 \\
 0  & 1   & 0 & 0 & 0 \\
 0 & 0 &-1 & 0 & 0 & 0 \\
 0 & 0 & 0 & g_{23}  & 0   & 0 \\
 0 & 0 & 0 & 0    &g_{23} \sin{(\theta)}^2  & 0 \\
 0 & 0 & 0 & 0 & 0 & g_{23}\cos{(\theta)}^2 \\
\end{array}
\right),
\end{align}
viz,
 \begin{equation}
 ds^2=(-dt^2+dx^2+dy^2) + g_{23}(d\theta^2 + \sin{\theta}^2 d\phi_{+}^2 + \cos{\theta}^2 d\phi_{-}^2).
 \end{equation}
 For the MN case, the effective Lagrangian density  is
\begin{eqnarray}
{\mathcal{L}^{MN}}_{\lambda^2} &=-& \frac{\sqrt{-g (k^2+N^2)}}{64 N \pi^5 {\alpha '}^3}\left( (\nabla^a \delta X^2) \nabla_a \delta X^2 +( (\nabla^a \delta\rho) \nabla_a \delta \rho  + m_\rho^2 \delta \rho^2) -\frac{1}{2}\delta F^{a b} \delta F_{a b} \right)\nonumber \\
&-& \frac{N}{16\pi^3} \sin{(2\theta)} \epsilon^{i j k} \delta A_i \partial_j \delta A_k,
\end{eqnarray}
where the indices $(a,b)$ span the full D5 coordinates,  $(x,y, t, \theta,\phi_+, \phi_-), $ while  $(i, j, k)$ span only the $\mathcal{M}^3$  component described by $(x, y, t)$.
With the exception of the massless field $\delta X^2$ and the relative sign between the  Chern-Simons term and the Yang-Mills term, the MNa effective Lagrangian are the same
\begin{eqnarray}
{\mathcal{L}^{MNa}}_{\lambda^2} &=-& \frac{\sqrt{-g (k^2+N^2)}}{64 N \pi^5 {\alpha '}^3}\left( ( (\nabla^a \delta\rho) \nabla_a \delta \rho  + m_\rho^2 \delta \rho^2) -\frac{1}{2}\delta F^{a b} \delta F_{a b} \right)\nonumber \\
&+& \frac{N}{16\pi^3} \sin{(2\theta)} \epsilon^{i j k} \delta A_i \partial_j \delta A_k.
\end{eqnarray}
 Due to the Chern-Simons term, both Lagrangians are gauge invariant only up to gauge parameters, $\Lambda$, that vanish on the boundary of $R^2\times \mathcal{I}$.

 The field equations for the MN case are:
\begin{equation}
\nabla^a \nabla_a \delta X^2 =0, \quad \nabla^a \nabla_a \delta \rho - {m_{\rho}}^2 \delta \rho =0, \quad
\nabla^a \delta F_{a b}= 4 \pi J^{\text{CS}}_b.\label{gaugeMN}
\end{equation}
Here the effective mass of the $\delta \rho$ field is
${m_\rho}^2= e^{-\Phi_0} \frac{8 k^2 +13 N^2}{9 N^3 \alpha '}$ and the current $J^{\text{CS}}_a$  arising from  the Chern-Simons term is
\begin{equation}
J^{\text{CS}}{}^i=2\, e^{-\Phi_0} \sqrt{(k^2 +N^2)} \,\frac{\alpha'}{N}\epsilon^{i j k} \partial_j \delta A_{k}, \label{CScurrent}
\end{equation}
for the $R^2\times \mathcal{I}$ coordinates and zero otherwise.

Similarly the MNa field equations are
\begin{eqnarray}
&\nabla^a \nabla_a \delta \rho - {m_{\rho}}^2 \delta \rho =0,\\
& \nabla^a \delta F_{a b}= - 4 \pi J^{\text{CS}}_b. \label{gaugeMNa}\end{eqnarray}
where here the renormalized mass of the $\delta \rho$ field is
${m_\rho}^2= e^{-\Phi_0} \frac{ 28 k^2 + N^2 (32 + \pi^2)}{48 N \alpha' (k^2 + N^2)}$ and the current  $J^{\text{CS}}_a$  is the same as the  MN case.   We will give an explicit solution in the appendix. A full discussion of the quadratic fluctuations will require more work. For the purpose of determining the L\"uscher term it will suffice to make qualitative arguments for the relevant modes of propagation that enter into the L\"uscher term.  We make those arguments presently.

\subsection{Massless Modes and  L\"uscher Term}
The  L\"uscher term is the long range, Coulombic contribution to the potential that arises from quantum corrections in the $k$-string.  Therefore the massless modes are the only modes that can contribute to the      L\"uscher term.  In the case of  the D5-brane described here, the geometry of the minimized manifold contains a  3-sphere so will carry a Laplacian that suggests that it is convenient to expand the fields in terms of  hyperspherical harmonics functions,  $\mathcal{T}^{(\kappa,m_{-},m_{+})}(\theta_,\,\phi_+,\,\phi_-)$.  Since we are interested in the massless modes of this theory, we expect them to correspond to the lowest lying modes of the 3-sphere.   One way to quickly extract this information is to consider the fields to be dependent only of the $(x, t)$ parameters on the full D5-brane.  Then we may integrate out the angular and $y$ dependence  in the field equations to recover the effective 2-D equations of motion.   This is similar  to integrating the angular variables to build an effective Lagrangians as in~\cite{Doran:2009pp}. Thus we can get an effective 2D $k$-string, and seek the number of massless modes in that part of the Lagrangian that is  quadratic in the fluctuations.  These $2D$ quadratic Lagrangians, up to total derivatives, are
for MN,

\begin{eqnarray}
   {\mathcal{L}^{MN}}_{2D} &\propto & \bigg[ c_X^{MN} \nabla_a \delta X^2 \nabla^a \delta X^2 + c_F^{MN} \delta F_{ab} \delta F^{ab} + c_\theta^{MN} \nabla_a \delta A_{\theta} \nabla^a \delta A_{\theta}\\
   &+&c_y^{MN} \nabla_a \delta A_{y} \nabla^a \delta A_{y}+ c_\phi^{MN} \nabla \widetilde{\delta A}_{\phi_i} \nabla^a \widetilde{\delta A}_{\phi_i} + c_\rho^{MN}(\nabla_a \delta \rho \nabla^a \delta \rho + m_\rho^2 \delta \rho^2) \bigg], \nonumber
\end{eqnarray}
 and
for MNa,\begin{align}
   {\mathcal{L}^{MNa}}_{2D} \propto \bigg[ &  c_F^{MNa} \delta F_{ab} \delta F^{ab} + c_\theta^{MNa} \nabla_a \widetilde{\delta A}_{\theta } \nabla^a \widetilde{\delta A}_{\theta} ++c_y^{MNa} \,\nabla_a \delta A_{y} \nabla^a \delta A_{y} \cr
   &+ c_\phi^{MNa} \nabla \widetilde{\delta A}_{\phi_i} \nabla^a \widetilde{\delta A}_{\phi_i}
    + c_\rho^{MNa}(\nabla_a \delta \rho \nabla^a \delta \rho + m_\rho^2 \delta \rho^2)\bigg],
\end{align}
where in both cases $i=+,-$ sums over the residual scalars from the original full $6$ component gauge field of the parent $6D$ theory. In these Lagrangians, the various $c^{MN}$'s and $c^{MNa}$'s are constants. The metric is $2D$ Minkowski, and the tilded fields are the field strength renormalizations of the untilded fields, for instance:
\begin{align}
    \widetilde{\delta A}_{\theta_i} =& Z^i \delta A_{\theta_i} ~~~\mbox{(no sum)}.
\end{align}

Also, $m_\rho^2$ is the effective mass of the $\delta \rho$ field, given by
\begin{align}
    m_\rho^2 = & \left\{ \begin{array}{l l}
                          \frac{ 8 k^2 + 13 N^2}{9 N \alpha'(k^2 + N^2)} & MN \\
                          \frac{ 28 k^2 + N^2 (32 + \pi^2)}{48 N \alpha' (k^2 + N^2)} & MNa
                        \end{array}
        \right. .
\end{align}
From the Lagrangians, we count six massless modes for MN: one from the $2D$ gauge field which has two degrees of freedom, minus the Gauss law constraint, and five massless scalars $\delta A_{y}$, $\delta A_{\theta}$, $\widetilde{\delta A}_{\phi_i}$, and $\delta X^2$. For MNa, we have the same counting, minus the $\delta X^2$ field. From an analysis parallel to that of our previous ones \cite{PandoZayas:2008hw,Doran:2009pp}, we conclude that the L\"uscher term fits that of our previous formulas
\begin{align}
   \delta E = - \frac{ (d + p -3 ) \pi}{24 L} + \beta
\end{align}
where $\beta$ is a renormalizeable constant.  This lends  credence to our belief that these solutions form a universality class for $k$-strings. It is interesting to note that this formula, which keeps appearing in all of our analyses of branes acting as $k$-strings, differs from L\"uscher's formula~\cite{Luscher:1980ac,Luscher:1980fr} by a number of degrees of freedom $p+1$ that is precisely the dimension of the D$p$-brane world volume.

\subsection*{L\"uscher term universality}

More explicitly, we find from all our previous brane analyses~\cite{PandoZayas:2008hw,Doran:2009pp,Stiffler:2009ma,Stiffler:2010pz} as well as the current one, the succinct formula for the L\"uscher term

\begin{equation}\label{eq:LuscherGG}
   V_{\mbox{L\"uscher}} = - \frac{(d+p -3) \pi}{24 L}
\end{equation}
where $d$ is the spatial dimension of the field theory and $p$ is the spatial dimension of the corresponding D$p$ brane realizing the configuration. It is worth emphasizing that the above formula is valid in the large $L$ length limit of the $k$-string.

This is in contrast to the formula which fits the lattice data well~\cite{Bringoltz:2008nd,Athenodorou:2010cs} at large $L$:
\begin{equation}\label{eq:LuscherLGT}
  V_{\mbox{L\"uscher}} = - \frac{(d - 2)\pi}{6L}~~~.
\end{equation}
However, clearly, by a judicious choice of $p$, one can acquire the same numerical value for the L\"uscher term that is found in lattice gauge theory
\begin{equation}
  p = 3d - 5,
\end{equation}
and this condition was satisfied in the classical analysis original due to Herzog~\cite{Herzog:2002ss} from which we later proved the form of the L\"uscher potential with the quantum corrections we computed in~\cite{Doran:2009pp}.

It is important to note that the lattice calculations are done with flux tubes that are tori~\cite{Bringoltz:2008nd, Athenodorou:2010cs}. As our D$p$ brane world volumes are not tori along the length of the $k$-string direction, in our regularization procedure~\cite{PandoZayas:2008hw,Doran:2009pp,Stiffler:2009ma,Stiffler:2010pz}, we chose periodic boundary conditions. This also makes it easier to extract results from the regularization procedure. Are the $k$-string representations of D-branes in a larger universality class of which the lattice gauge theory $SU(N)$ results are a subset? This is an interesting question which we hope to pursue more in the future.

Let us finish our discussion of the L\"uscher term by confronting the expectations from field theory to the results of holography. First, in field theory it is expected that the $k$-string L\"uscher term be independent of $k$. Our result presented in equation (\ref{eq:LuscherGG}) is, indeed, independent of $k$. The holographic formula has a $p$ dependence which could be troubling to the field theorist as its interpretation as a field theory parameter is not immediate. Note, however, that holographically $p$ is present on very general grounds and has to do with the counting of massless excitations above the classical configurations, those are precisely the quantum fluctuations that contribute to the L\"uscher term. Another way of seeing this term from the field theory point of view is simply to think of it as shifting the effective dimension where the excitations of the confining $k$-string live.

\section{Conclusions}\label{sec:conclusion}

In this paper we have considered $k$-string configurations in strongly coupled, confining field theories by studying their dual D5-brane configurations. We have also computed the L\"uscher term which requires a one-loop computation on the D-brane side. We have applied some of the  developments of the AdS/CFT correspondence in its conformal realization \cite{Gomis:2006sb,Gomis:2006im} to clarify the question of the precise representations described by the D-brane configurations.

Arguably, our main result is summarized in table  (\ref{tab:comparetensions}) where we have obtained qualitative agreement with other approaches to the problem of tension of $k$-strings such as the Hamiltonian approach in $2+1$ pioneered by Karabali and Nair and the lattice approach as articulated by Teper and collaborators \cite{Bringoltz:2006gp,Athenodorou:2010cs}. It is worth noting that the holographic approach provides answers that await for comparison with other methods. For example, the lattice approach has not yet arrived at a conclusion in the case of some $k$-strings in the totally symmetric representation. Analogously, the Hamiltonian approach of Karabali-Nair is not developed enough to produce a results in 3+1 dimensional field theories. Less we forget that the answer of holographic methods requires a large $N$ limit, namely, $N\to \infty$ with $k/N$ held fixed. All in all, the interaction of various approaches to the tensions of $k$-strings seems to be a fertile ground for cross-field fertilization.

One important result of our calculations is a universal formula for the L\"uscher term in holographic models of $k$ strings. It will be interesting to check our formula against lattice calculations now that improved computational methods allow for precision calculation of the L\"uscher term.

\section*{Acknowledgements}

We thank Adi Armoni for incisive questions on the first version of the manuscript. We are also thankful to Parameswaran Nair for various clarifications and correspondence.
 V.G.J.R. acknowledges the hospitality of the Michigan Center for Theoretical Physics. L.A.P.Z. is thankful to the Aspen Center for Physics
for hospitality.  This research was supported in part by the National Science Foundation under Grants Nos. 1066293 (Aspen) and 1067889 (University of Iowa) by Department
of Energy under grant DE-FG02-95ER40899 (University of Michigan) and by the endowment of the John S.~Toll Professorship, the University of Maryland
 Center for String \& Particle Theory, National Science Foundation Grant PHY-0354401.

\appendix

\section{Explicit solution}

Here we exhibit examples of explicit solutions to the k-string fluctuations.  This solution  displays some peculiarities associated with the geometry one of which is the apparent spin $\frac{3 }{2}$  modes locked on the $S^3$.
\subsection{Spin Zero Fields}
In both the MN and MNa cases, the spin-zero bosonic fields satisfy the field equations given by the d'Alembertian on $\mathcal{M}^3 =R^2\times \mathcal{I}\times S^3$.
\begin{equation}
\nabla^a \nabla_a \Upsilon\  - {M}^2 \,\Upsilon\ =0,
\end{equation}
where $M$ can be zero.  For spin zero fields, the d'Alembertian decouples into a d'Alembertian operator on $\mathcal{M}^3$ and the Laplacian  operator on $S^3$
$$
\nabla^2= \nabla_{\mathcal{M}^3}^2+ \frac{1}{\exp{(-\Phi_0)} N  \alpha'}\nabla_{\mathcal{S}^3}^2.
$$
Then a suitable ansatz for $\Upsilon$  is
\begin{equation}
\Upsilon(x,y,t, \theta, \phi_+, \phi_-)=\Upsilon_{\mathcal{M}^3}(x,y,t) \Upsilon_{S^3}( \theta, \phi_+, \phi_- ).
\end{equation}
The eigenstate for $\nabla_{\mathcal{M}^3}$ operator may be written as
\begin{equation}
\Upsilon^{p_x,p_y,\omega}_{\mathcal{M}^3}(x,y,t)= \exp{(ip_x + ip_y y -i\omega \,t)} \Upsilon_0,
\end{equation}
while  $\Upsilon_{S^3}( \theta, \phi_+, \phi_- )$ are eigenstates of $\nabla_{\mathcal{S}^3}^2 $  the hyperspherical harmonics $\mathcal{T}^{(\kappa,m_{-},m_{+})}(\theta_,\,\phi_+,\,\phi_-)$ satisfying
\begin{equation}
\nabla_{\mathcal{S}^3}^2  \mathcal{T}^{(\kappa,m_{-},m_{+})}(\theta_,\,\phi_+,\,\phi_-)= -\kappa (2+\kappa)\mathcal{T}^{(\kappa,m_{-},m_{+})}(\theta_,\,\phi_+,\,\phi_-)
\end{equation}
for positive integers $\kappa$, and $m_+,m_-= -\frac{k}{2},\cdots,\frac{k}{2}$.  The have $(\kappa +1)^2$  degeneracy for a given $\kappa$.
With this we explicitly write that
\begin{equation}
\Upsilon(x,y,t, \theta, \phi_+, \phi_-)=\sum_{} C^{\, p_x,p_y,p_z}\,\Upsilon^{p_x,p_y,\omega}_{\mathcal{M}^3}(x,y,t)\Upsilon^{j\,(\kappa,m_+,m_-)}_{S^3}( \theta, \phi_+, \phi_- ),
\end{equation}
with the two hyperspherical harmonics corresponding to $j=1,2$ given by,
\begin{align}
\Upsilon^{1\, (\kappa,m_+,m_-)}_{S^3}( \theta, \phi_+, \phi_- )=e^{(i m_+ \phi_+ +i m_- \phi)}\,
 i^{-(m_++m_-)} \cos^{-(m_++m_-)}\times\nonumber\\ {}_2F_1(-(\frac{\kappa}{2} + \frac{m_+}{2}+ \frac{m_-}{2});(\frac{\kappa}{2} - \frac{m_+}{2}- \frac{m_-}{2});1 - m_+- m_-;\cos^2(\theta)),\\ \Upsilon^{2\,(\kappa,m_+,m_-)}_{S^3}( \theta, \phi_+, \phi_- )=e^{(i m_+ \phi_+ +i m_- \phi)}\,
 i^{(m_++m_-)} \cos^{(m_++m_-)}\times\nonumber\\ {}_2F_1(-(\frac{\kappa}{2} - \frac{m_+}{2}- \frac{m_-}{2});(\frac{\kappa}{2} + \frac{m_+}{2}+ \frac{m_-}{2}+1;1 + m_++ m_-;\cos^2(\theta)).\end{align}
The ${}_2F_1$ are hypergeometric functions of the second kind and the energies, $\omega$, are given by
\begin{equation}
\omega= \pm \frac{\sqrt{e^{-\Phi_0}(\kappa(2+\kappa) + e^{\Phi_0}(p_x^2+p_y^2+M^2)N \alpha')}}{N \alpha'}.
\end{equation}

\subsection{Vector Bosons }

The field equations Eqs.[\ref{gaugeMNa},\ref{gaugeMN}], are identical for both the MN and MNa up to a sign in the CS current,
\begin{equation}
\nabla^a F_{a b} = \nabla^a\nabla_a \delta A_b - \nabla_b \nabla^a \delta A_a -R_{c b} \,\delta A^c = 4 \pi J^{\text{CS}}_b.
\end{equation}
Now the Ricci tensor is  $$R^{\text{\,effective}}_{a b}= \frac{2 \exp{(-\Phi_0})}{N \alpha'} g_{a b} \,\,\,\,\text{for} \,\,\,\,\,(a,b)=(\theta, \phi_+, \phi_-) $$ and zero otherwise.  Furthermore ${J^{\text{CS}}}_a$ is non-trivial only in the $\mathcal{M}^3$ component.  The Lorentz gauge, $\nabla^a \delta A_a=0$ ,  provides a convenient gauge choice as the field components $(\delta A_x, \delta A_y, \delta A_t)$ decouple from $(\delta A_\theta, \delta A_{\phi_+}, \delta A_{\phi_-})$.   Below we give two solutions.  The second solution is interesting because it exhibits an interesting spin $\frac{3}{2}$ behavior on the $S^3$ component.

\subsubsection{Solution 1}

In this solution, five of the fields propagate while $A_\theta=0$.    We write
\begin{align}
A_x(\xi)=&{A_x}^0\, p_x\, e^{i(-\omega_0 t + p_x x+p_y y+m_0 \phi_- + n_0 + \phi_+)} a_x(\theta)\cr
A_y(\xi)=&{A_x}^0\, p_y\,  e^{i(-\omega_0 t + p_x x+p_y y+m_0 \phi_- + n_0 + \phi_+)}a_x(\theta)\cr
A_t(\xi)=-&{A_x}^0\,\omega\, e^{i(-\omega_0 t + p_x x+p_y y+m_0 \phi_- + n_0 + \phi_+)}a_x(\theta)\cr
A_{\phi_+}(\xi)=&{A_+}^0\,\, e^{i(-\omega_+ t + q_x x+q_y y+m_1 \phi_- )}a_{\phi_+}(\theta)\cr
A_{\phi_-}(\xi)=&{A_-}^0\,\, e^{i(-\omega_- t + r_x x+r_y y+n_1 \phi_+ )}a_{\phi_-}(\theta),\cr
\end{align}
where
\begin{align}
\omega_+ =& \pm\sqrt{\frac{e^{-\Phi_0} \kappa_+^2+q_x^2 N \alpha' +q_y^2 N \alpha'}{N \alpha'}}\cr
\omega_- =& \pm\sqrt{\frac{e^{-\Phi_0} \kappa_-^2+r_x^2 N \alpha' +r_y^2 N \alpha'}{N \alpha'}},
\end{align}
with $\kappa_\pm$ integers.  Here $\vec{p}$, $\vec{q}$ and $\vec{r}$ are momenta on $\mathcal{M}^3$. The $\theta$  dependent fields satisfy hypergeometric differential equations are are given by
\small{
\begin{eqnarray}
a_x(\theta)& = &\frac{\sec (\theta ) \cos ^2(\theta)^{\frac{1}{2}-\frac{m}{2}}}{{\sqrt{\sin ^2(\theta)}}}
\left(-\sin^2(\theta )\right)^{\frac{1}{2}-\frac{n}{2}}
   \left(c_2 (-1)^m \cos^2(\theta )^m  {}_2F_1\left(\frac{m-n}{2},\frac{1}{2}
   (m-n+2);m+1;\cos^2(\theta )\right) \right.  \nonumber \\
   &+&\left. c_1 \,    {}_2F_1\left(\frac{1}{2} (-m-n),\frac{1}{2}
   (-m-n+2);1-m;\cos^2(\theta)\right)\right),
\end{eqnarray}
}

\begin{eqnarray}
a_{\phi_-}(\theta) &=&c_1 i^{-n_1} \sin ^2(\theta
   )^{-\frac{n_1}{2}} \, _2F_1\left(-\frac{\kappa
   _-}{2}-\frac{n_1}{2},\frac{\kappa
   _-}{2}-\frac{n_1}{2};1-n_1;\sin ^2(\theta
   )\right)\\
   &+ & c_2 i^{n_1} \sin ^2(\theta
   )^{\frac{n_1}{2}} \,
   _2F_1\left(\frac{n_1}{2}-\frac{\kappa
   _-}{2},\frac{\kappa
   _-}{2}+\frac{n_1}{2};n_1+1;\sin ^2(\theta
   )\right),
\end{eqnarray}
and
\begin{eqnarray}
a_{\phi_+}(\theta)&=&\, c_1 i^{-m_1} \cos ^{-m_1}(\theta ) \,
   _2F_1\left(-\frac{\kappa
   _+}{2}-\frac{m_1}{2},\frac{\kappa
   _+}{2}-\frac{m_1}{2};1-m_1;\cos ^2(\theta
   )\right)\\
   &+& c_2 i^{m_1} \cos ^{m_1}(\theta ) \,
   _2F_1\left(\frac{m_1}{2}-\frac{\kappa
   _+}{2},\frac{\kappa
   _+}{2}+\frac{m_1}{2};m_1+1;\cos ^2(\theta
   )\right).
\end{eqnarray}
\subsubsection{Solution 2}

There are four cases that constitute these solutions where $A_\theta$ is non-trivial on the $S^3$,  These solutions are determined by the parameters $(s_1, s_2)$.  The four cases correspond to the four pairs for  $(s_1,s_2)$ given by $(1,1), (1,-1), (-1,1),$ and $ (-1,-1). $ We  write this covariantly divergence free solution as:
 \begin{equation}
 A_a(\xi)=\sum C(p_x,p_y,\omega,m_+,m_-,\kappa) A^C_a(\xi),
 \end{equation}
 where the $C$'s are constants and the components of $A^C_a$ are
\begin{align}
A_x(\xi; p_{x},p_{y},\omega, m_+,m_-)=& i^{1 - m_+} p_x A_x^0  \,  e^{-i(p_x x + p_y y -\omega \, t +m_+ \phi_+ +m_-\phi_- ) }\times\nonumber\\ &\sin{(\theta)}^{-m_+} \mathcal{W}(m_+,m_-,\theta )\\
A_y(\xi; p_{x},p_{y},\omega, m_+,m_-)=& i^{1 - m_+} p_y A_x^0  \,  e^{-i(p_x x + p_y y -\omega \, t +m_+ \phi_+ +m_-\phi_- ) }\times\nonumber\\ &\sin{(\theta)}^{-m_+} \mathcal{W}(m_+,m_-,\theta )\\
A_t(\xi; p_{x},p_{y},\omega, m_+,m_-)=&-  i^{1 - m_+} \omega  A_x^0  \, e^{-i(p_x x + p_y y -\omega \, t +m_+ \phi_+ +m_-\phi_- ) }\times\nonumber\\  &\sin{(\theta)}^{-m_+} \mathcal{W}(m_+,m_-,\theta )\\
A_\theta(\xi; p'_{x},p'_{y},\omega')=s_1&  \frac{3}{2} A_{\phi_+}^0 e^{i( p'_x x + p'_y -  \frac{1}{3}\omega' t +\frac{2}{3}\,s_2 (\phi_+ + s_{1}\phi_{ -})}\frac{\sin{(2 \theta)}^{\frac{1}{3}}}{1-\cos{(4 \theta)}}\\
A_{\phi_+}(\xi; p'_{x},p'_{y},\omega')=& \frac{1}{2} A_{\phi_+}^0  e^{i( p'_x x + p'_yy -  \frac{1}{3}\omega' t)}\,\mathcal{\:Q}_1(\theta,s_1,s_2 )\\
A_{\phi_-}(\xi; p_{x},p_{y},\omega)=& - A_{\phi_+}^0 e^{i( p'_x x + p'_y -  \frac{1}{3}\omega' t)}\,\mathcal{\:Q}_2(\theta, s_1, s_2).
\end{align}
With,
\begin{align}
\mathcal{W}(m_+,m_-,\theta )=&\nonumber\\(-1)^{1-m_+}\cos{(\theta)}^{-m_-} \big(C_1 \:\:{}_2 F_{1}&\bigg(\frac{-m_--m_+ }{2},\frac{2-m_--m_+}{2};1-m_-;\cos{(\theta)}^2\bigg) + \nonumber\\ (-1)^{m_-}\; C_2 \sin{(\theta)}^{-m_+}\cos{(\theta)}^{2 m_-} \:\:{}_2 F_{1}&\bigg(\frac{m_--m_+ }{2},\frac{2+m_--m_+}{2};1+m_-;\cos{(\theta)}^2\bigg) \big),
\end{align}
and the frequencies are given through,$$\;\omega=\pm\sqrt{p_x^2 +p_y^2}, \;\; \;\omega'=\pm \sqrt{\frac{\frac{4e^{-\Phi_0}}{9}+((p'_x)^2+(p'_y)^2)N\alpha'}{N \alpha'}}.$$ Two Meijer functions are contained in the $\mathcal{\:Q}_1(\theta )$ and $\mathcal{\:Q}_2(\theta )$. Explicitly these are

\begin{align}
\mathcal{\:Q}_1(\theta )= 2\, c_2\, G_{2,2}^{2,0}\left(\cos^2(\theta )|
\begin{array}{c}
 \frac{2}{3},\frac{4}{3} \\
 0,0 \\
\end{array}
\right)
+2\, c_1 \; _2F_1\left(-\frac{1}{3},\frac{1}{3};1;\cos^2(\theta )\right)\nonumber\\
- s_1  s_2 3\, i\,{\sin (2 \theta )}^{\frac{1}{3}} \cot (\theta ) e^{\frac{2}{3} i (\phi_-
 +s_1 \phi_+)},\label{spin321}
\end{align}
\begin{align}
\mathcal{\:Q}_2(\theta )= 2\, c_3\, G_{2,2}^{2,0}\left(\cos ^2(\theta )|
\begin{array}{c}
 \frac{2}{3},\frac{4}{3} \\
 0,1 \\
\end{array}
\right)-2 A_{\phi_-} c_4 \cos ^2(\theta ) \,
   _2F_1\left(\frac{2}{3},\frac{4}{3};2;\cos ^2(\theta )\right)\nonumber \\+s_2\,3\, i A_{\phi_+}
   {\sin (2 \theta )}^{\frac{1}{3}} \tan (\theta ) e^{\frac{2}{3} i (\phi_- +s_1\,\phi_+)}.\label{spin322}
\end{align}
Notice that in Eqs.[\ref{spin321},\ref{spin322}] that the $\phi$ angles must be rotated by $3 \pi$ in order to return to their initial value. This suggests that this solution has spin $\frac{3}{2}$ features on the $S^3$ component.

\bibliographystyle{JHEP}
\bibliography{bibliography}

\providecommand{\href}[2]{#2}\begingroup\raggedright\begin{thebibliography}{10}

\bibitem{Maldacena:1997re}
J.~M. Maldacena, {\it {The large N limit of superconformal field theories and
  supergravity}},  {\em Adv. Theor. Math. Phys.} {\bf 2} (1998) 231--252,
  [\href{http://xxx.lanl.gov/abs/hep-th/9711200}{{\tt hep-th/9711200}}].

\bibitem{Witten:1998qj}
E.~Witten, {\it {Anti-de Sitter space and holography}},  {\em Adv. Theor. Math.
  Phys.} {\bf 2} (1998) 253--291,
  [\href{http://xxx.lanl.gov/abs/hep-th/9802150}{{\tt hep-th/9802150}}].

\bibitem{Gubser:1998bc}
S.~S. Gubser, I.~R. Klebanov, and A.~M. Polyakov, {\it {Gauge theory
  correlators from non-critical string theory}},  {\em Phys. Lett.} {\bf B428}
  (1998) 105--114, [\href{http://xxx.lanl.gov/abs/hep-th/9802109}{{\tt
  hep-th/9802109}}].

\bibitem{Aharony:1999ti}
O.~Aharony, S.~S. Gubser, J.~M. Maldacena, H.~Ooguri, and Y.~Oz, {\it {Large N
  field theories, string theory and gravity}},  {\em Phys. Rept.} {\bf 323}
  (2000) 183--386, [\href{http://xxx.lanl.gov/abs/hep-th/9905111}{{\tt
  hep-th/9905111}}].

\bibitem{Klebanov:2000hb}
I.~R. Klebanov and M.~J. Strassler, {\it {Supergravity and a confining gauge
  theory: Duality cascades and chiSB-resolution of naked singularities}},  {\em
  JHEP} {\bf 08} (2000) 052,
  [\href{http://xxx.lanl.gov/abs/hep-th/0007191}{{\tt hep-th/0007191}}].

\bibitem{Maldacena:2000yy}
J.~M. Maldacena and C.~Nunez, {\it {Towards the large N limit of pure N = 1
  super Yang Mills}},  {\em Phys. Rev. Lett.} {\bf 86} (2001) 588--591,
  [\href{http://xxx.lanl.gov/abs/hep-th/0008001}{{\tt hep-th/0008001}}].

\bibitem{Chamseddine:2001hk}
A.~H. Chamseddine and M.~S. Volkov, {\it {Non-Abelian vacua in D = 5, N = 4
  gauged supergravity}},  {\em JHEP} {\bf 04} (2001) 023,
  [\href{http://xxx.lanl.gov/abs/hep-th/0101202}{{\tt hep-th/0101202}}].

\bibitem{Maldacena:2001pb}
J.~M. Maldacena and H.~S. Nastase, {\it {The supergravity dual of a theory with
  dynamical supersymmetry breaking}},  {\em JHEP} {\bf 09} (2001) 024,
  [\href{http://xxx.lanl.gov/abs/hep-th/0105049}{{\tt hep-th/0105049}}].

\bibitem{Cvetic:2001ma}
M.~Cvetic, G.~W. Gibbons, H.~Lu, and C.~N. Pope, {\it {Supersymmetric
  non-singular fractional D2-branes and NS-NS 2-branes}},  {\em Nucl. Phys.}
  {\bf B606} (2001) 18--44, [\href{http://xxx.lanl.gov/abs/hep-th/0101096}{{\tt
  hep-th/0101096}}].

\bibitem{Shifman:2005eb}
M.~Shifman, {\it {k strings from various perspectives: QCD, lattices, string
  theory and toy models}},  {\em Acta Phys. Polon.} {\bf B36} (2005)
  3805--3836, [\href{http://xxx.lanl.gov/abs/hep-ph/0510098}{{\tt
  hep-ph/0510098}}].

\bibitem{Herzog:2001fq}
C.~P. Herzog and I.~R. Klebanov, {\it {On string tensions in supersymmetric
  SU(M) gauge theory}},  {\em Phys.Lett.} {\bf B526} (2002) 388--392,
  [\href{http://xxx.lanl.gov/abs/hep-th/0111078}{{\tt hep-th/0111078}}].

\bibitem{Herzog:2002ss}
C.~P. Herzog, {\it {String tensions and three dimensional confining gauge
  theories}},  {\em Phys. Rev.} {\bf D66} (2002) 065009,
  [\href{http://xxx.lanl.gov/abs/hep-th/0205064}{{\tt hep-th/0205064}}].

\bibitem{PandoZayas:2008hw}
L.~A. Pando~Zayas, V.~G. Rodgers, and K.~Stiffler, {\it {Luscher Term for
  k-string Potential from Holographic One Loop Corrections}},  {\em JHEP} {\bf
  0812} (2008) 036, [\href{http://xxx.lanl.gov/abs/0809.4119}{{\tt
  arXiv:0809.4119}}].

\bibitem{Doran:2009pp}
C.~A. Doran, L.~A. Pando~Zayas, V.~G. Rodgers, and K.~Stiffler, {\it {Tensions
  and Luscher Terms for (2+1)-dimensional k-strings from Holographic Models}},
  {\em JHEP} {\bf 0911} (2009) 064,
  [\href{http://xxx.lanl.gov/abs/0907.1331}{{\tt arXiv:0907.1331}}].

\bibitem{Stiffler:2009ma}
K.~Stiffler, {\it {Mesons From String Theory}},
  \href{http://xxx.lanl.gov/abs/0909.5681}{{\tt arXiv:0909.5681}}.

\bibitem{Stiffler:2010pz}
K.~M. Stiffler, {\it {A Walk Through Superstring Theory With an Application to
  Yang-Mills Theory: K-strings and D-branes as Gauge/Gravity Dual Objects}},
  \href{http://xxx.lanl.gov/abs/1012.0021}{{\tt arXiv:1012.0021}}.

\bibitem{Drukker:2005kx}
N.~Drukker and B.~Fiol, {\it {All-genus calculation of Wilson loops using
  D-branes}},  {\em JHEP} {\bf 0502} (2005) 010,
  [\href{http://xxx.lanl.gov/abs/hep-th/0501109}{{\tt hep-th/0501109}}].

\bibitem{Gomis:2006sb}
J.~Gomis and F.~Passerini, {\it {Holographic Wilson loops}},  {\em JHEP} {\bf
  08} (2006) 074, [\href{http://xxx.lanl.gov/abs/hep-th/0604007}{{\tt
  hep-th/0604007}}].

\bibitem{Gomis:2006im}
J.~Gomis and F.~Passerini, {\it {Wilson loops as D3-branes}},  {\em JHEP} {\bf
  01} (2007) 097, [\href{http://xxx.lanl.gov/abs/hep-th/0612022}{{\tt
  hep-th/0612022}}].

\bibitem{Faraggi:2011bb}
A.~Faraggi and L.~A. Pando~Zayas, {\it {The Spectrum of Excitations of
  Holographic Wilson Loops}},  {\em JHEP} {\bf 1105} (2011) 018,
  [\href{http://xxx.lanl.gov/abs/1101.5145}{{\tt arXiv:1101.5145}}].

\bibitem{Faraggi:2011ge}
A.~Faraggi, W.~Mueck, and L.~A. Pando~Zayas, {\it {One-loop Effective Action of
  the Holographic Antisymmetric Wilson Loop}},  {\em Phys.Rev.} {\bf D85}
  (2012) 106015, [\href{http://xxx.lanl.gov/abs/1112.5028}{{\tt
  arXiv:1112.5028}}].

\bibitem{Lucini:2001nv}
B.~Lucini and M.~Teper, {\it {Confining strings in SU(N) gauge theories}},
  {\em Phys. Rev.} {\bf D64} (2001) 105019,
  [\href{http://xxx.lanl.gov/abs/hep-lat/0107007}{{\tt hep-lat/0107007}}].

\bibitem{Lucini:2002wg}
B.~Lucini and M.~Teper, {\it {SU(N) gauge theories in (2+1)-dimensions: Further
  results}},  {\em Phys.Rev.} {\bf D66} (2002) 097502,
  [\href{http://xxx.lanl.gov/abs/hep-lat/0206027}{{\tt hep-lat/0206027}}].

\bibitem{Bringoltz:2006gp}
B.~Bringoltz and M.~Teper, {\it {String tensions of SU(N) gauge theories in 2+1
  dimensions}},  {\em PoS} {\bf LAT2006} (2006) 041,
  [\href{http://xxx.lanl.gov/abs/hep-lat/0610035}{{\tt hep-lat/0610035}}].

\bibitem{Karabali:2007mr}
D.~Karabali and V.~P. Nair, {\it {The robustness of the vacuum wave function
  and other matters for Yang-Mills theory}},  {\em Phys. Rev.} {\bf D77} (2008)
  025014, [\href{http://xxx.lanl.gov/abs/0705.2898}{{\tt arXiv:0705.2898}}].

\bibitem{Kinar:1998vq}
Y.~Kinar, E.~Schreiber, and J.~Sonnenschein, {\it {Q anti-Q potential from
  strings in curved space-time: Classical results}},  {\em Nucl.Phys.} {\bf
  B566} (2000) 103--125, [\href{http://xxx.lanl.gov/abs/hep-th/9811192}{{\tt
  hep-th/9811192}}].

\bibitem{Brandhuber:1998er}
A.~Brandhuber, N.~Itzhaki, J.~Sonnenschein, and S.~Yankielowicz, {\it {Wilson
  loops, confinement, and phase transitions in large N gauge theories from
  supergravity}},  {\em JHEP} {\bf 9806} (1998) 001,
  [\href{http://xxx.lanl.gov/abs/hep-th/9803263}{{\tt hep-th/9803263}}].

\bibitem{Kruczenski:2004me}
M.~Kruczenski, L.~A. Pando~Zayas, J.~Sonnenschein, and D.~Vaman, {\it {Regge
  trajectories for mesons in the holographic dual of large-N(c) QCD}},  {\em
  JHEP} {\bf 0506} (2005) 046,
  [\href{http://xxx.lanl.gov/abs/hep-th/0410035}{{\tt hep-th/0410035}}].

\bibitem{Pawelczyk:2000hy}
J.~Pawelczyk and S.-J. Rey, {\it {Ramond-ramond flux stabilization of
  D-branes}},  {\em Phys.Lett.} {\bf B493} (2000) 395--401,
  [\href{http://xxx.lanl.gov/abs/hep-th/0007154}{{\tt hep-th/0007154}}].

\bibitem{Camino:2001at}
J.~Camino, A.~Paredes, and A.~Ramallo, {\it {Stable wrapped branes}},  {\em
  JHEP} {\bf 0105} (2001) 011,
  [\href{http://xxx.lanl.gov/abs/hep-th/0104082}{{\tt hep-th/0104082}}].

\bibitem{Craps:1999nc}
B.~Craps, J.~Gomis, D.~Mateos, and A.~Van~Proeyen, {\it {BPS solutions of a
  D5-brane world volume in a D3-brane background from superalgebras}},  {\em
  JHEP} {\bf 9904} (1999) 004,
  [\href{http://xxx.lanl.gov/abs/hep-th/9901060}{{\tt hep-th/9901060}}].

\bibitem{Camino:1999xx}
J.~Camino, A.~Ramallo, and J.~Sanchez~de Santos, {\it {World volume dynamics of
  D-branes in a D-brane background}},  {\em Nucl.Phys.} {\bf B562} (1999)
  103--132, [\href{http://xxx.lanl.gov/abs/hep-th/9905118}{{\tt
  hep-th/9905118}}].

\bibitem{Athenodorou:2010cs}
A.~Athenodorou, B.~Bringoltz, and M.~Teper, {\it {Closed flux tubes and their
  string description in D=3+1 SU(N) gauge theories}},  {\em JHEP} {\bf 1102}
  (2011) 030, [\href{http://xxx.lanl.gov/abs/1007.4720}{{\tt
  arXiv:1007.4720}}].

\bibitem{Douglas:1995nw}
M.~R. Douglas and S.~H. Shenker, {\it {Dynamics of SU(N) supersymmetric gauge
  theory}},  {\em Nucl.Phys.} {\bf B447} (1995) 271--296,
  [\href{http://xxx.lanl.gov/abs/hep-th/9503163}{{\tt hep-th/9503163}}].

\bibitem{Armoni:2003ji}
A.~Armoni and M.~Shifman, {\it {On k string tensions and domain walls in N=1
  gluodynamics}},  {\em Nucl.Phys.} {\bf B664} (2003) 233--246,
  [\href{http://xxx.lanl.gov/abs/hep-th/0304127}{{\tt hep-th/0304127}}].

\bibitem{Agarwal:2007ns}
A.~Agarwal, D.~Karabali, and V.~Nair, {\it {Yang-Mills theory in 2+1
  dimensions: Coupling of matter fields and string-breaking effects}},  {\em
  Nucl.Phys.} {\bf B790} (2008) 216--239,
  [\href{http://xxx.lanl.gov/abs/0705.0394}{{\tt arXiv:0705.0394}}].

\bibitem{Agarwal:2009gb}
A.~Agarwal, {\it {Mass Deformation of Super Yang-Mills in D=2+1}},  {\em
  Phys.Rev.} {\bf D80} (2009) 105020,
  [\href{http://xxx.lanl.gov/abs/0909.0959}{{\tt arXiv:0909.0959}}].

\bibitem{Agarwal:2012bn}
A.~Agarwal and V.~Nair, {\it {Supersymmetry and Mass Gap in 2+1 Dimensions: A
  Gauge Invariant Hamiltonian Analysis}},  {\em Phys.Rev.} {\bf D85} (2012)
  085011, [\href{http://xxx.lanl.gov/abs/1201.6609}{{\tt arXiv:1201.6609}}].

\bibitem{Armoni:2011dw}
A.~Armoni, D.~Dorigoni, and G.~Veneziano, {\it {k-String Tension from
  Eguchi-Kawai Reduction}},  {\em JHEP} {\bf 1110} (2011) 086,
  [\href{http://xxx.lanl.gov/abs/1108.6196}{{\tt arXiv:1108.6196}}].

\bibitem{Polchinski:1998rr}
J.~Polchinski, {\it {String theory. Vol. 2: Superstring theory and beyond}}, .

\bibitem{Armoni.Talk:2010}
A.~Armoni, {\it {Everything you wanted to know about k-strings but you were
  afraid to ask }},  {\em ECT, Trento} {\bf
  http://pyweb.swan.ac.uk/~pyarmoni/trento.pdf} (2010).

\bibitem{Luscher:1980ac}
M.~Luscher, {\it {Symmetry Breaking Aspects of the Roughening Transition in
  Gauge Theories}},  {\em Nucl. Phys.} {\bf B180} (1981) 317.

\bibitem{Luscher:1980fr}
M.~Luscher, K.~Symanzik, and P.~Weisz, {\it {Anomalies of the Free Loop Wave
  Equation in the WKB Approximation}},  {\em Nucl. Phys.} {\bf B173} (1980)
  365.

\bibitem{Bringoltz:2008nd}
B.~Bringoltz and M.~Teper, {\it {Closed k-strings in SU(N) gauge theories : 2+1
  dimensions}},  {\em Phys. Lett.} {\bf B663} (2008) 429--437,
  [\href{http://xxx.lanl.gov/abs/0802.1490}{{\tt arXiv:0802.1490}}].

\end{thebibliography}\endgroup

\end{document}